\documentclass[12pt]{article}
\usepackage[cp1251]{inputenc}
\usepackage[english]{babel}

\begin{document}

\begin{center}
{\Large  Method for searching higher symmetries for quad graph equations}

\vskip 0.2cm

{Rustem  N. Garifullin}\footnote{e-mail:  rustem@matem.anrb.ru}\\

{Ufa Institute of Mathematics, Russian Academy of Science,\\
Chernyshevskii Str., 112, Ufa, 450008, Russia}\\

\bigskip

{Elena V. Gudkova}\footnote{e-mail: elena.gudkova79@mail.ru}

{Department of Applied Mathematics and Mechanics,\\ Ufa State Petroleum Technical University,
 \\Kosmonavtov str., 1, Ufa, 450062, Russia }\\

\bigskip
{Ismagil T. Habibullin}\footnote{e-mail: habibullinismagil@gmail.com}\\

{Ufa Institute of Mathematics, Russian Academy of Science,\\
Chernyshevskii Str., 112, Ufa, 450008, Russia}\\

\bigskip


\end{center}
\begin{abstract}
Generalized symmetry integrability test for discrete equations on the square lattice is studied. Integrability conditions are discussed. A method for searching higher symmetries (including non-autonomous ones) for quad graph equations is suggested based on characteristic vector fields.
\end{abstract}
\bigskip

{\it Keywords:} higher symmetry, non-autonomous symmetry, quad-graph equation, classification, characteristic vector field, integrability
conditions.

\section{Introduction}

Consider nonlinear equations on the quad graph (or double discrete chains) of the form
\begin{equation}\label{ddhyp}
u_{1,1}=f(u,u_1,  \bar u_{1} ).
\end{equation}
Here the unknown $u=u(m,n)$ is a function of two discrete variables $m,n$. For the sake of convenience we use the following notations: $u_k=u(m+k,n)$, $\bar{u}_k=u(m,n+k)$, $u_{1,1}=u(m+1,n+1)$.
Function $f$ is supposed to be locally analytic, it depends essentially on all three arguments. In other words equation (\ref{ddhyp}) can be rewritten in any of the following forms
\begin{equation}
u_{i,j}=f^{i,j}(u,u_i, \bar u_{j}),\label{i,j}\quad \mbox{with}\quad i=\pm1,\,j=\pm1.
\end{equation}

Equations of the form (\ref{ddhyp}) have important applications and the problem of the complete description of their integrable cases is challenging. At present there are various approaches for studying integrability of discrete models. In \cite{NijhoffWalker}, \cite{BobenkoSuris}, \cite{Nijhoff}, \cite{abs}, \cite{Hietarinta} the property of consistency around a cube is considered as a criterion of integrability. Another classification method called vanishing of the algebraic entropy is developed in \cite{BellonViallet}, \cite{NijhoffRamani}, \cite{GKP}, \cite{HydonViallet}. The symmetry approach is approved to provide a powerful classification tool for integrable discrete and continuous models \cite{hydon}, \cite{YL1}, \cite{YL2}, \cite{Xenitidis}, \cite{RasinHydon}, \cite{Mikhailov}, \cite{tongas}.

It was observed earlier that characteristic vector fields provide an effective classification tool for integrable models either in the case of Darboux integrability \cite{Shabat}, \cite{HPZh1}, \cite{HPZh2} or S-integrability \cite{ZhiberMur}, \cite{HG}, \cite{HG2}. Recall that in the discrete case these vector fields are defined as follows $Y_1=\bar D^{-1}\frac{\partial}{\partial \bar u_1}\bar D$ and $Y_{-1}=\bar D\frac{\partial}{\partial \bar u_{-1}}\bar D^{-1}$ and have the following coordinate representation \cite{hab}
\begin{equation}\label{defY1}
Y_1=\frac{\partial}{\partial u}+x\frac{\partial}{\partial  u_{1}}+\frac{1}{x_{-1}}\frac{\partial}{\partial  u_{-1}}+
xx_1\frac{\partial}{\partial  u_{2}}+\frac{1}{x_{-1}x_{-2}}\frac{\partial}{\partial u_{-2}}+ ...,
\end{equation} and
\begin{equation}\label{defY-1}
Y_{-1}=\frac{\partial}{\partial u}+y\frac{\partial}{\partial  u_{1}}+\frac{1}{y_{-1}}\frac{\partial}{\partial  u_{-1}}+
yy_1\frac{\partial}{\partial  u_{2}}+\frac{1}{y_{-1}y_{-2}}\frac{\partial}{\partial u_{-2}}+ ...,
\end{equation}
where $x=\bar D^{-1}(\frac{\partial f(u,u_1,\bar u_1)}{\partial \bar u_{1}})=-\frac{\partial f^{1,-1}(u,u_1,\bar u_{-1})/\partial u}{\partial f^{1,-1}(u,u_1,\bar u_{-1})/\partial u_{1}}$ and $y=\bar D(\frac{\partial f^{1,-1}(u,u_1,\bar u_{-1})}{\partial \bar u_{-1}})=-\frac{\partial f(u,u_1,\bar u_{1})/\partial u}{\partial f(u,u_1,\bar u_{1})/\partial u_{1}}$. 
Here the shift operators $D$ and $\bar D$ act due to the rule $Dh(m,n)=h(m+1,n),$ $\bar Dh(m,n)=h(m,n+1)$.
Vector fields $Y_{1}$, $Y_{-1}$, $X_{1}=\frac{\partial}{\partial \bar u_{-1}}$ and $X_{-1}=\frac{\partial}{\partial \bar u_{1}}$ considered as operators applied to the variables $\bar u_1, \bar u_{-1}, u, u_{\pm1},u_{\pm2},...$ satisfy the following conjugation relations
\begin{equation}\label{autXY}
DX_1D^{-1}=pX_1, \, DY_1D^{-1}=\frac{1}{x}Y_1,\,DX_{-1}D^{-1}=qX_{-1}, \, DY_{-1}D^{-1}=\frac{1}{y}Y_{-1},
\end{equation} 
where $p=D(\frac{\partial f^{-1,-1}(u,u_{-1},\bar u_{-1})}{\partial \bar u_{-1}})=\frac{1}{\partial f^{1,-1}(u,u_1,\bar u_{-1})/\partial \bar u_{-1}}$ and $q=D(\frac{\partial f^{-1,1}(u,u_{-1},\bar u_{1})}{\partial \bar u_{1}})=\frac{1}{\partial f(u,u_1,\bar u_{-1})/\partial \bar u_{1}}$. 

To stress the close connection of these operators with the symmetry approach we note that restrictions of the characteristic vector fields $Y_1$ and $Y_{-1}$ on the set of functions depending only on $u,u_1,u_{-1}$ coincide with the differential operators,
introduced by P.E.Hydon \cite{hydon} and successfully applied in \cite{RasinHydon}, \cite{YL1}, \cite{YL2} to look for conservation laws and symmetries
\begin{equation}\label{defY1a}
A=\frac{\partial}{\partial u}-\frac{f^{1,-1}_u}{f^{1,-1}_{u_1}}\frac{\partial}{\partial  u_{1}}-\frac{f^{-1,-1}_u}{f^{-1,-1}_{u_{-1}}}\frac{\partial}{\partial  u_{-1}},
\end{equation} and
\begin{equation}\label{defY-1a}
B=\frac{\partial}{\partial u}-\frac{f_u}{f_{u_1}}\frac{\partial}{\partial  u_{1}}-\frac{f^{-1,1}_u}{f^{-1,1}_{u_{-1}}}\frac{\partial}{\partial  u_{-1}}.
\end{equation}
To compare operators (\ref{defY1}), (\ref{defY-1}), (\ref{defY1a}), (\ref{defY-1a}) use the following formulas
\begin{equation}\label{toshow}
1/x_{-1}=-f^{-1,-1}_u/f^{-1,-1}_{u_{-1}}\quad \mbox{and}\quad 1/y_{-1}=-f^{-1,1}_u/f^{-1,1}_{u_{-1}}. 
\end{equation}

Our work was stimulated by the results obtained by Levi and Yamilov. In the over mentioned articles \cite{YL1}, \cite{YL2} they made a good progress in a formalization of the theory of integrable discrete models and gave an effective tool to look for autonomous symmetries of (\ref{ddhyp}) of the form 
\begin{equation}\label{YLsym}
u_t=g(u,u_1,u_{-1},\bar u_1,\bar u_{-1}).
\end{equation}

The goal of the present article is to suggest a method suitable for searching higher symmetries of equation (\ref{ddhyp}) of general form (\ref{sym}) depending also on discrete arguments $m$, $n$. It is important since there are integrable equations (\ref{ddhyp}) having no any symmetry of the form (\ref{YLsym}) but having symmetries (\ref{sym}). We know at least two examples of such a kind, one of them is equation (\ref{introductionHV}) considered below and the other is Adler's  discretization of the Tzitzeica equation (AdT) recently suggested in \cite{adler11}.

In \cite{Mikhailov} an alternative approach to discrete equations is studied. Necessary conditions of integrability are found under assumption that equation (\ref{ddhyp}) admits a recursion operator given as a formal series of a special form
$${\bf R}=r_1D+r+r_{-1}D^{-1}+r_{-2}D^{-2}+\dots \, ,$$
with coefficients $r_{j}$.

The article is organized as follows. In section 2 we discuss on a sequence of integrability conditions implied by the requirement of existence of symmetries of sufficiently high order. These integrability conditions are derived by consecutive differentiation of the linearized equation with respect to the dynamical variables. They are in fact functional equations. We give the first four of the conditions in an explicit form  (see Proposition 1 below). Among them two were deduced earlier in \cite{YL1}. It makes a sense to compare these integrability conditions with those derived in \cite{Mikhailov}. It is an intriguing question, whether AdT equation satisfies the integrability conditions from \cite{Mikhailov}. In the next section 3 by using characteristic vector fields we deduce differential consequences of the integrability conditions from section 2. The consequences are given as systems of linear first order PDE. Compatibility conditions of these systems are effective necessary integrability conditions since they are easily checked for any given quad graph equation of the form (\ref{ddhyp}). At the same time the systems provide an algorithm of constructing symmetries. In Lemma2 it is proved that any symmetry (\ref{sym}) below splits down into a sum of two functions like $g(m,n,u_j,u_{j-1},...,u_{j'}, \bar u_{k}, \bar u_{k-1},...\bar u_{k'} )=F(m,n,u_j,u_{j-1},...,u_{j'})+G(m,n,\bar u_{k}, \bar u_{k-1},...\bar u_{k'})$. Recall that in the particular case (\ref{YLsym}) this fact has been proved in \cite{RH}. Efficiency of the symmetry finding algorithm is demonstrated in \S\S4-5. In the fourth section a non-autonomous symmetry 
\begin{equation}\label{nonautosym}
g=(-1)^{m+n}C\frac{u(u^2-1)(u_1u_{-1}+1)}{(uu_1+u_1-u+1)(uu_{-1}-u_{-1}+u+1)} 
\end{equation}
is found for the equation
\begin{equation}\label{introductionHV}
u_{1,1}u(u_{1}-1)(\bar u_{1}+1)+(u_{1}+1)(\bar u_{1}-1)=0
\end{equation}
suggested in \cite{HydonViallet}. In a recent paper \cite{Gramm} the connection of this equation with discrete SG equation is discussed.
In section 5 a fifth order symmetry $u_t=g$ is evaluated for the same equation (\ref{introductionHV}), where
\begin{eqnarray}
&&g={\frac {uu_{{1}} \left( u_{{1}}-1 \right)  \left( u^2-1
 \right)   }{ \left( u_{{1}}+1+u_{{1}}u-
u \right) ^{2} \left( 1-u_{{1}}+u_{{2}}+u_{{2}}u_{{1}} \right) }
}+\nonumber \\ 
&&-{\frac {u_{{-1}}u \left( u_{{-1}}+1 \right)   \left( u^2-1 \right) }{ \left( -u_{{-1}}+u_{{-1}}u
+1+u \right) ^{2} \left( 1+u_{{-1}}-u_{{-2}}+u_{{-2}}u_{{-1}}
 \right) }}-
\label{introductionHVsym}\\
&& -{\frac {u \left( u^2-1 \right)  
 \left( 1+u_{{1}}u_{{-1}} \right)  \left( uu_{{-1}}-u_{{-1}}-u u_{{1}}-u_{{1}} \right) }{ \left( u_{{1}}+1+uu_{{1}}-u
 \right) ^{2} \left( 1-u_{{-1}}+uu_{{-1}}+u \right) ^{2}}
}.\nonumber 
\end{eqnarray}

For some certain cases the systems (\ref{s}), (\ref{sv}), (\ref{econd3}), (\ref{econd4}) derived in section 3 are sufficient to find the final form of a fifth order symmetry if it does exist. It is the case for the equation (\ref{introductionHV}) mentioned above. However, they are not the only differential consequences of the integrability conditions listed in Proposition 1. In section 6 we briefly discuss how to derive additional systems of linear first order PDE consistency of which is also necessary for existence of higher symmetries.

\section{Higher symmetries and integrability conditions}

Existence of higher symmetries is believed to be an important indication of integrability. 
Assume that chain (\ref{ddhyp}) admits a higher symmetry (generally non-autonomous) of the form
\begin{equation}\label{sym}
u_{t}=g(m,n,u_j,u_{j-1},...,u_{j'}, \bar u_{k}, \bar u_{k-1},...\bar u_{k'} ),
\end{equation}
where $j\geq1$, $j'\leq-1$, $k\geq1$, $k'\leq-1$.
As it is well known the function $g$ should satisfy the linearization of chain (\ref{ddhyp})
\begin{equation}\label{linearisation}
D\bar{D}g=f_{u_1}Dg+f_{\bar u_1}\bar Dg+f_ug,
\end{equation}

{\bf Lemma 1.} Linearized equation  (\ref{linearisation}) can be rewritten in any of the following equivalent forms
\begin{eqnarray}\label{linearisation-1,1}
D^{-1}\bar{D}g&=&f_{u_{-1}}^{-1,1}D^{-1}g+f_{\bar u_1}^{-1,1}\bar Dg+f_u^{-1,1}g,\\
D\bar{D}^{-1}g&=&f_{u_{1}}^{1,-1}Dg+f_{\bar u_{-1}}^{1,-1}\bar D^{-1}g+f_u^{1,-1}g,\nonumber \\
D^{-1}\bar{D}^{-1}g&=&f_{u_{-1}}^{-1,-1}D^{-1}g+f_{\bar u_{-1}}^{-1,-1}\bar D^{-1}g+f_u^{-1,-1}g.\nonumber 
\end{eqnarray}
{\bf Proof.} Let us prove the first statement of Lemma 1. To this end apply the operator $D^{-1}$ to equation (\ref{linearisation}) and simplify it due to the relations
$$D^{-1}(f_u/f_{\bar u_1})=-f^{-1,1}_{u_{-1}}, \quad D^{-1}(1/f_{\bar u_1})=f^{-1,1}_{\bar u_{1}}, \quad D^{-1}(f_{u_1}/f_{\bar u_1})=-f^{-1,1}_{u}.$$
The other formulas are verified in a similar way.
 
 Differentiate equation  (\ref{linearisation}) with respect to the highest order variable $u_{j+1}$ and get (see \cite{YL1})
\begin{equation}\label{differentiated}
D\bar{D}g_{u_j}D^jf_{u_1}=f_{u_1}Dg_{u_j}.
\end{equation}
Set $z:=\log g_{u_j}$ and rewrite the last equation in the form of a conservation law
\begin{equation}\label{z}
\bar{z}_{1}=z+(D^{-1}-D^{j-1})\log f_{u_1}.
\end{equation}
Differentiation of (\ref{linearisation-1,1}) with respect to the lowest order variable $u_{j'-1}$ yields
\begin{equation}\label{j'differentiated}
D^{-1}\bar{D}g_{u_{j'}}D^{j'}f_{u_{-1}}^{-1,1}=f_{u_{-1}}^{-1,1}Dg_{u_{j'}}.
\end{equation}
Obviously,  equation (\ref{j'differentiated}) implies for $v:=\log g_{u_{j'}}$
\begin{equation}\label{v}
\bar{v}_{1}=v+(D-D^{j'+1})\log f_{u_{-1}}^{-1,1}.
\end{equation}

Actually, equations (\ref{z}) and (\ref{v}) provide necessary conditions of integrability of equation (\ref{ddhyp}): there exist numbers $j$ and $j'$ such that functions $(D^{-1}-D^{j-1})\log f_{u_1}$ and $(D-D^{j'+1})\log f_{u_{-1}}^{-1,1}$ belong to the image of the operator $E-\bar D$, where $E$ is the identity operator. These integrability conditions derived for the first time in \cite{YL1} are highly nonlocal, since each equation contains the values of unknowns $z,v$ taken at two different values of their arguments $z=z(m,n,u_j,u_{j-1},...,u_{j'}, \bar u_{k}, \bar u_{k-1},...\bar u_{k'})$, $v=v(m,n,u_j,u_{j-1},...,u_{j'}, \bar u_{k}, \bar u_{k-1},...\bar u_{k'})$ and $\bar z=z(m,n+1,u_{j,1},u_{j-1,1},...,u_{j',1}, \bar u_{k+1}, \bar u_{k},...\bar u_{k'+1})$, $\bar v=v(m,n+1,u_{j,1},u_{j-1,1},...,u_{j',1}, \bar u_{k+1}, \bar u_{k},...\bar u_{k'+1})$. In other words equations (\ref{z}) and (\ref{v}) are functional equations. 

To derive the next necessary condition of existence of a symmetry of the form (\ref{sym}) with $j\geq2$ differentiate equation (\ref{linearisation}) with respect to the variable $u_j$
\begin{eqnarray}\label{differentiated2}
D\bar{D}(g_{u_{j-1}})D^{j-1}(f_{u_1})+D\bar{D}(g_{u_j})[D^j(f_{u})+D^j(f_{\bar u_1})D^{j-1}(f_{u_1})]=f_{u_1}D(g_{u_{j-1}})+\\
+f_{\bar u_1}\bar D(g_{u_j})D^{j-1}(f_{u_1})+f_ug_{u_j}.\nonumber
\end{eqnarray}
Put $z^{(1)}:=g_{u_{j-1}}$ and rewrite equation (\ref{differentiated2}) as follows:
\begin{equation}\label{z1}
\bar z_1^{(1)}=r(m,n,\bar u_1,u_{j},u_{j-1},...u_{-1})z^{(1)}+R(m,n,\bar u_1,u_{j},u_{j-1},...u_{j'}),
\end{equation}
where $r=\frac{D^{-1}(f_{u_1})}{D^{j-2}(f_{u_1})}$ and $$R=\frac{1}{D^{j-2}(f_{u_1})}D^{-1}\left\{f_{\bar u_1}\bar D(g_{u_j})D^{j-1}(f_{u_1})+f_ug_{u_j}- D\bar{D}(g_{u_j})D^j(f_{u})\right\}-\bar D(g_{u_j})D^{j-1}(f_{\bar u_1}).$$
Thus a symmetry exists only when function $R$ is in the image of the operator $\bar D-rE$. 

Differentiating equation (\ref{linearisation-1,1}) with respect to the variable $u_{j'}$ one obtains for any $j'\leq-2$
\begin{eqnarray}
D^{-1}\bar{D}(g_{u_{j'+1}})D^{j'+1}(f_{u_{-1}}^{-1,1})+D^{-1}\bar{D}(g_{u_j'})[D^{j'}(f_{u}^{-1,1})+D^{j'}(f_{\bar u_{1}}^{-1,1})D^{j'+1}(f_{u_{-1}}^{-1,1})]= \label{differentiated3}\\
=f_{u_{-1}}^{-1,1}D^{-1}(g_{u_{j'+1}})+f_{\bar u_1}^{-1,1}\bar D(g_{u_{j'}})D^{j'+1}(f_{u_{-1}}^{-1,1})+f_u^{-1,1}g_{u_{j'}} .\nonumber
\end{eqnarray}
Put $v^{(1)}:=g_{u_{j'+1}}$ and rewrite the last equation as follows
\begin{equation}\label{v1}
\bar v_1^{(1)}=\widetilde r(m,n,\bar u_1,u_{1},u,...u_{j'+1})v^{(1)}+\widetilde R(m,n,\bar u_1,u_{j},u_{j-1},...u_{j'}),
\end{equation}
where the functions 
$$\widetilde r=\frac{D(f^{-1,1}_{u_{-1}})}{D^{j'+2}(f^{-1,1}_{u_{-1}})}$$ 
and 
$$\widetilde R=\frac{1}{D^{j'+2}(f^{-1,1}_{u_{-1}})}D\left\{f^{-1,1}_{\bar u_1}\bar D(g_{u_{j'}})D^{j'+1}(f^{-1,1}_{u_{-1}})+f^{-1,1}_ug_{u_{j'}}-D^{-1}\bar{D}(g_{u_{j'}})D^{j
'}(f^{-1,1}_{u})\right\}-$$ $$-\bar D(g_{u_{j'}})D^{j'+1}(f^{-1,1}_{\bar u_1})$$ 
are expressed through known quantities $f^{-1,1}$, $v$, their derivatives and shifts. Obviously equation (\ref{v1}) provides one more integrability condition: function $\widetilde R$ should be in the image of the operator  $\bar D-\widetilde rE$. 

Finalizing the reasonings above we obtain the following statement.

{\bf Proposition 1}. If an equation of the form (\ref{ddhyp}) admits a higher symmetry of sufficiently great order then the conditions hold for some entire $j$ and $j'$:\\
\noindent
1) $(D^{-1}-D^{j-1})\log f_{u_1}\in Im (E-\bar D)$ for $j\geq 1$ and $j'\leq-1$ ;\\
2) $(D-D^{j'+1})\log f_{u_1}^{-1,1}\in Im (E-\bar D)$ for $j\geq 1$ and $j'\leq-1$; \\
3) $R(m,n,\bar u_1,u_{j+1},u_{j},...u_{j'})\in Im (rE-\bar D)$ for $j\geq 2$ and $j'\leq-2$;\\
4) $\widetilde R(m,n,\bar u_1,u_{j+1},u_{j},...u_{j'})\in Im (\widetilde rE-\bar D)$ for $j\geq 2$ and $j'\leq-2.$

In such a way one can derive a sequence of integrability conditions given as functional equations. Obviously, in this form integrability conditions are not very effective. Below in the next section, we suggest a rule to derive some differential consequences of these equations which have a form of systems of first order partial differential equations and can effectively be studied by standard methods of differential algebra.

\section{Algorithm of finding symmetries}

Concentrate first on equation (\ref{z}). Note that $z$ might depend only on the variables $u_j,u_{j-1},...,u_{j'}$, it does not depend on $\bar u_j$ with $j\neq0$. Indeed, suppose that $z$ depends on $\bar u_k,\, k>0$ and does not depend on $\bar u_s,\, s>k$, then evidently $\bar z_1$  depends on $\bar u_{k+1}$ that is not possible since it contradicts equation (\ref{z}). The case $k<0$ is studied analogously. Reduce (\ref{z}) to a system of the first order PDE's. To this end apply the operator $\bar D^{-1}\frac{\partial}{\partial \bar u_1}$ to both sides of the equation and have
\begin{equation}\label{Y1z}
Y_{1}z=f^{(1)},
\end{equation}
where $f^{(1)}=\bar D^{-1}\frac{\partial}{\partial \bar u_1}(D^{-1}-D^{j-1})\log f_{u_1}$ and $Y_1=\bar D^{-1}\frac{\partial}{\partial \bar u_1}\bar D$ (see formula (\ref{defY1}) above). Note that the coefficients of equation (\ref{Y1z}) depend on the variable $\bar u_{-1}$, in spite of the solution $z$ cannot depend on it. That is why one has to put an additional condition
\begin{equation}\label{X1}
X_{1}z=0,
\end{equation}
where $X_1=\frac{\partial}{\partial \bar u_{-1}}$. 

In a similar way the functional equation (\ref{z}) implies two more differential equations. Indeed, the equation can be represented as  
$\bar{z}_{-1}=z-\bar D^{-1}(D^{-1}-D^{j-1})\log f_{u_1}$. Applying the operator $\bar DX_1$ gives
\begin{equation}\label{Y-1z}
Y_{-1}z=f^{(-1)},
\end{equation}
where $f^{(-1)}=-Y_{-1}[(D^{-1}-D^{j-1})\log f_{u_1}]$ and $Y_{-1}=\bar D\frac{\partial}{\partial \bar u_{-1}}\bar D^{-1}$ (see formula (\ref{defY-1}) above).

Thus we come up to a system of the first order non-homogeneous linear equations 
\begin{eqnarray}
X_{1}z&=&0, \nonumber\\
Y_{1}z&=&f^{(1)},\label{s}\\
X_{-1}z&=&0, \nonumber\\
Y_{-1}z&=&f^{(-1)}. \nonumber
\end{eqnarray}
Note that for $j=1$, $j'=-1$ system (\ref{s}) in essence coincides with that suggested in \cite{YL1}. 
Compatibility of system (\ref{s}) is necessary for existence of symmetry (\ref{sym}). Emphasize that generally system (\ref{s}) is not closed. To close it we have to add all its linearly independent differential consequences obtained by taking cross applications of the operators such as $[X_1,Y_1]z=X_1f^{(1)}$, $[Y_1,Y_{-1}]z=Y_1f^{(-1)}-Y_{-1}f^{(1)}$ etc. In such a way we arrive at a system of the form 
\begin{eqnarray}\label{cs1}
L_sz:&=&\sum^{j}_{i=j'}a_{s,i}\frac{\partial z}{\partial u_{i}}=F^{(s)},\quad s=1,2,...N,\\
X_{1}z&=&0, \nonumber\\
X_{-1}z&=&0, \nonumber
\end{eqnarray}
satisfying the condition: any further cross application of the operators $L_s$ gives an equation of the form $[L_s,L_r]z=L_s(F^{(r)})-L_r(F^{(s)})$ linearly expressed through already known equations (\ref{cs1}). Then the system is closed.

Introduce notations $p^{(i)}=\frac{\partial z}{\partial u_{i}}$ and rewrite system (\ref{cs1}) as a system of linear algebraic equations 
\begin{equation}\label{cs2}
\sum^{j}_{i=j'}a_{s,i}p^{(i)}=F^{(s)},\quad s=1,2,...N.
\end{equation}
Two last  equations in (\ref{cs1}) are valid automatically since $z$ does not depend on $\bar u_1, \bar u_{-1}$.
Due to the well known Kronecker–Capelli theorem, system of linear equations (\ref{cs2}) is compatible if and only if the rank of the coefficient matrix $A=(a_{s,i})$ is equal to that of the augmented matrix $B$ obtained from $A$ by adding the column of free terms $F^{(s)}$.  Thus the condition $rank(A)=rank(B)$ is necessary for existence of a symmetry (\ref{sym}).

Let us deduce some differential consequences of equation (\ref{v}). Just applying the reasonings above to this equation we obtain
\begin{eqnarray}
X_{1}v&=&0, \nonumber\\
Y_{1}v&=&\widetilde f^{(1)},\label{sv}\\
X_{-1}v&=&0, \nonumber\\
Y_{-1}v&=&\widetilde f^{(-1)}, \nonumber
\end{eqnarray}
where $\widetilde f^{(1)}=\bar D^{-1}\frac{\partial}{\partial \bar u_1}(D-D^{j'+1})\log f_{u_{-1}}^{-1,1}$ and 
$\widetilde f^{(-1)}=-Y_{-1}[(D-D^{j'+1})\log f_{u_{-1}}^{-1,1}]$. Similarly to the previous case one can close system (\ref{sv}) and then reduce it to a system of linear algebraic equations. We omit this part.

Study the third condition of Proposition 1, which requires the consistency of the functional equation (\ref{z1}). Note that again unknown $z^{(1)}$  does not depend on the variable $\bar u_1$, hence applying the operator  
$\bar D^{-1}\frac{\partial}{\partial\bar u_1}$ to equation (\ref{z1}) yields
\begin{equation}\label{y22}
Y_1z^{(1)}=\bar D^{-1}(r_{\bar u_1})\bar z_{-1}^{(1)}+\bar D^{-1}R_{\bar u_1}.
\end{equation}
Exclude $\bar z_{-1}^{(1)}$ by means of the same equation (\ref{z1}) to obtain
\begin{equation}\label{y222}
Y_1z^{(1)}=A^{(1)}z^{(1)}+B^{(1)},
\end{equation}
where $A^{(1)}=\bar D^{-1}(\log r)_{\bar u_{1}}$ and $B^{(1)}=\bar D^{-1}(R_{\bar u_{1}}-R(\log r)_{\bar u_{1}})$. 
Rewrite equation (\ref{z1}) as follows 
\begin{equation}\label{z11}
z^{(1)}=\frac{1}{r}(\bar z_1^{(1)}-R)
\end{equation}
and shift it back by applying the operator $\bar D^{-1}$
\begin{equation}\label{z1-1}
\bar z_{-1}^{(1)}=\bar D^{-1}(\frac{1}{r})z^{(1)}-\bar D^{-1}(\frac{R}{r}).
\end{equation}
Apply the operator $\bar D\frac{\partial}{\partial \bar u_{-1}}$ to the last equation to get
\begin{equation}\label{z1Y-1}
Y_{-1}z^{(1)}=Y_{-1}(\frac{1}{r})\bar z_1^{(1)}-Y_{-1}(\frac{R}{r}),
\end{equation}
which obviously is rewritten as follows 
\begin{equation}\label{y222-1}
Y_{-1}z^{(1)}=A^{(-1)}z^{(1)}+B^{(-1)},
\end{equation}
where $A^{-1}=rY_{-1}(1/r)$ and $B^{-1}=-\frac{1}{r}Y_{-1}(R)$. As it was done above we put two more equations 
$X_{1}z^{(1)}=0$ and $X_{-1}z^{(1)}=0$. Thus we find a consequence of the necessary condition 3) of the Proposition 1. The following system of the first order linear PDE should be consistent
\begin{eqnarray}
&&Y_1z^{(1)}=A^{(1)}z^{(1)}+B^{(1)},\nonumber \\
&&X_{1}z^{(1)}=0,\nonumber \\
&&Y_{-1}z^{(1)}=A^{(-1)}z^{(1)}+B^{(-1)},\label{econd3} \\
&&X_{-1}z^{(1)}=0.\nonumber 
\end{eqnarray}

Deduce differential consequences from the fourth condition in Proposition 1, to this end apply the operator  
$\bar D^{-1}\frac{\partial}{\partial\bar u_1}$ to equation (\ref{v1}):
\begin{equation}\label{y22v}
Y_1v^{(1)}=\bar D^{-1}(\widetilde r_{\bar u_1})\bar v_{-1}^{(1)}+\bar D^{-1}\widetilde R_{\bar u_1}.
\end{equation}
Exclude $\bar v_{-1}^{(1)}$ by means of equation (\ref{v1}) to obtain
\begin{equation}\label{y222v}
Y_1\widetilde v^{(1)}=\widetilde A^{(1)}v^{(1)}+\widetilde B^{(1)},
\end{equation}
where $\widetilde A^{(1)}=\bar D^{-1}(\log \widetilde r)_{\bar u_{1}}$ and $\widetilde B^{(1)}=\bar D^{-1}(\widetilde R_{\bar u_{1}}-\widetilde R(\log \widetilde r)_{\bar u_{1}})$. 
Rewrite equation (\ref{v1}) as follows 
\begin{equation}\label{v11}
v^{(1)}=\frac{1}{\widetilde r}(\bar v_1^{(1)}-\widetilde R)
\end{equation}
and shift it back by applying the operator $\bar D^{-1}$
\begin{equation}\label{v1-1}
\bar v_{-1}^{(1)}=\bar D^{-1}(\frac{1}{\widetilde r})( v^{(1)}-\bar D^{-1}(\frac{\widetilde R}{\widetilde r}).
\end{equation}
Apply the operator $\bar D\frac{\partial}{\partial \bar u_{-1}}$ to the last equation to get
\begin{equation}\label{v1Y-1}
Y_{-1}v^{(1)}=Y_{-1}(\frac{1}{\widetilde r})\bar v_1^{(1)}-Y_{-1}(\frac{\widetilde R}{\widetilde r}),
\end{equation}
which obviously is rewritten as follows 
\begin{equation}\label{y222-1v}
Y_{-1}v^{(1)}=\widetilde A^{(-1)}v^{(1)}+\widetilde B^{(-1)},
\end{equation}
where $\widetilde A^{-1}=\widetilde rY_{-1}(1/\widetilde r)$ and $\widetilde B^{-1}=-\frac{1}{\widetilde r}Y_{-1}(\widetilde R)$. As it was done above we put two more equations 
$X_{1}v^{(1)}=0$ and $X_{-1}v^{(1)}=0$. Thus we find a consequence of the necessary condition 4) of the Proposition 1. The following system of the first order linear partial differential equations should be compatible
\begin{eqnarray}
&&Y_1v^{(1)}=A^{(1)}z^{(1)}+B^{(1)},\nonumber \\
&&X_{1}v^{(1)}=0,\nonumber \\
&&Y_{-1}v^{(1)}=A^{(-1)}v^{(1)}+B^{(-1)},\label{econd4} \\
&&X_{-1}v^{(1)}=0.\nonumber 
\end{eqnarray}

Continuing this way one can derive $i$-th integrability condition which is formulated as a compatibility conditions for a system of equations for the function $z^{(i)}:=g_{u_{s}}$ with $j'\leq s\leq j$. It is easy to prove by induction that all of the functions $g_{u_{s}}$ depend only on the variables $m,n,u_j,u_{j-1},...,u_{j'}$. This observation allows to prove the following statement on decomposition of a symmetry of the lattice (\ref{ddhyp}):

{\bf Lemma 2}. Any symmetry (\ref{sym}) of the lattice (\ref{ddhyp}) splits down into a sum of two functions $$g(m,n,u_j,u_{j-1},...,u_{j'}, \bar u_{k}, \bar u_{k-1},...\bar u_{k'} )=F(m,n,u_j,u_{j-1},...,u_{j'})+G(m,n,\bar u_{k}, \bar u_{k-1},...\bar u_{k'}).$$ Here it is supposed that $u=\bar u=u$.

{\bf Proof}. We observed above that $$g_{u_s}=a_s(m,n,u_j,u_{j-1},...,u_{j'}), \hbox{where }j'\leq s\leq j.$$ These formulas yield $$g_{u_s\bar u_r}\equiv 0\ \hbox{where }j'\leq s\leq j,\ k'\leq r \leq k,\  r\neq0,\ s\neq0.$$ The latter immediately proves the lemma.

\section{Evaluation of a non-autonomous symmetry for quad graph equation}

It is proved in \cite{YL2} that equation (\ref{introductionHV}) does not admit any autonomous symmetry of the form (\ref{YLsym}) but it is conjectured there that it might have a non-autonomous symmetry:
\begin{equation}\label{nonasym}
u_t=g(m,n,u_{-1},u,u_1).
\end{equation}
Apply the scheme above to equation (\ref{introductionHV}) to look for its symmetry (\ref{nonasym}). Function $z=\log g_{u_1}$
solves equation (\ref{z}) of the form $Y_1z=f^{(1)},\ Y_{-1}z=\widetilde{f}^{(1)}$. Rewrite it in coordinates:
$$\frac{u_{-1}^2-1}{2u}\frac{\partial z}{\partial u_{-1}}+\frac{\partial z}{\partial u}+\frac{2u_1}{u^2-1}\frac{\partial z}{\partial u_{1}}=\frac{u^2-2u-1}{u(u^2-1)};$$
$$\frac{2u_{-1}}{u^2-1}\frac{\partial z}{\partial u_{-1}}-\frac{\partial z}{\partial u}+\frac{u_{1}^2-1}{2u}\frac{\partial z}{\partial u_{1}} =-\frac{u^2u_1+2u^2-u_1}{u(u^2-1)}.$$
Note that the latter is not closed. To close this system one has to add one more equation $[Y_1,Y_{-1}]z=Y_1\widetilde{f}^{(1)}-Y_{-1}f^{(1)}$. Now solve the obtained system of three equations with respect to the partial derivatives of $z$:
$$\frac{\partial z}{\partial u_{-1}}=0,\quad \frac{\partial z}{\partial u_{1}}=-\frac{2(u+1)}{u_1(u+1)+1-u}, $$
$$\frac{\partial z}{\partial u}=-\frac{2(u_1-1)}{u_1(u+1)+1-u}+\frac1u+\frac1{u+1}+\frac1{u-1}. $$
This triple of equations is an overdetermined system of equations for $z$ which is consistent because the previous system is closed. 
Hence its general solution $z$ is easily found. It contains arbitrary function $C_1(m,n)$ depending on both discrete variables:
$$z=\log {\frac {C_1(m,n)u(u^2-1)}{(uu_1+u_1-u+1)^2}}$$
Substitution of $z$ to (\ref{z}) yields
$\log {\frac {-C_1(m,n+1)}{C_1(m,n)}}=0$. The last equation is easily solved $C_1(m,n)=(-1)^nC_2(m)$ where $C_2(m)$ is an arbitrary function of one discrete variable.
  
Therefore by solving equation $z=\log g_{u_1}$ we find:
$$g=-(-1)^n {\frac {C_2(m)u(u-1)}{uu_1+u_1-u+1} }+g_2(m,n,u_{-1},u).$$

For the further specification consider $v(u_{-1},u)=\log g_{u_{-1}}=\log g_{2,{u_{-1}}}$ which due to the general scheme solves the equations:
$$\frac{u_{-1}^2-1}{2u}\frac{\partial v}{\partial u_{-1}}+\frac{\partial v}{\partial u}=\frac{2u^2-2uu_{-1}+u_{-1}}{u(u^2-1)};$$
$$\frac{2u_{-1}}{u^2-1}\frac{\partial v}{\partial u_{-1}}-\frac{\partial v}{\partial u} =-\frac{1-u^2-2u}{u(u^2-1)}.$$
Now solve them with respect to the derivatives
$$\frac{\partial v}{\partial u_{-1}}=-\frac{2(u-1)}{u_{-1}(u-1)+u+1}, $$
$$\frac{\partial v}{\partial u}=-\frac{2(u_{-1}+1)}{u_{-1}(u-1)+1+u}+\frac1u+\frac1{u+1}+\frac1{u-1}$$
and then find $v$:
$$v=\log {\frac {C_3(m,n)u(u^2-1)}{(uu_{-1}-u_{-1}+u+1)^2}}.$$
Substitute it to the difference equation (\ref{v}) and obtain the relation
$$\log {\frac {-C_3(m,n+1)}{C_3(m,n)}}=0,$$  solution to which is $C_3(m,n)=(-1)^nC_4(m)$, where $C_4(m)$
is an arbitrary function of $m$. As a result function $g$ takes the form;
$$g=-(-1)^n {\frac {C_2(m)u(u-1)}{uu_1+u_1-u+1} }-(-1)^n {\frac {C_4(m)u(u+1)}{uu_{-1}-u_{-1}+u+1} }+g_3(u).$$

By substituting the function obtained into the linearized equation and applying to the result the following operator:
$$\frac{(u_{-1}-1)^2}2D^{-1}\frac\partial{\partial u_1}\frac1{u_1}\bar D^{-1}\frac{\partial}{\partial \bar u_1}\frac1{\bar u_1^2-1}$$ we come to the relation
$$g_3^{\prime\prime}(u)-\frac1ug_3^\prime(u)+\frac1{u^2}g_3(u)=\frac{u_{-1}^2+1}{(uu_{-1}-u_{-1}+u+1)^2}(C_2(m-1)-C_4(m))$$ which immediately implies:
$$g_3^{\prime\prime}(u)-\frac1ug_3^\prime(u)+\frac1{u^2}g_3(u)=0, \qquad C_4(m)=C_2(m-1). $$
Therefore $g_3(u)=C_5(m,n)u+C_6(m,n)u\log u,$ where the coefficients do not depend on $u$ but might depend on $m,n$. 

Now we have
$$g=-(-1)^n {\frac {C_2(m)u(u-1)}{uu_1+u_1-u+1} }-(-1)^n {\frac {C_2(m-1)u(u+1)}{uu_{-1}-u_{-1}+u+1} }+C_5(m,n)u+C_6(m,n)u\log u $$
Substitute it to the linearized equation and apply the operator
$$\frac1{u_1}\bar D^{-1}\frac{\partial}{\partial \bar u_1}\frac1{\bar u_1^2-1},$$ to get:
$$2(u^2+1)C_5(m,n)+2((u^2+1)\log u-u^2+1)C_6(m,n)+2(u^2-1)C_6(m+1,n)-$$ $$-C_2(m)(u-1)^2+C_2(m-1)(u+1)^2=0.$$ Collecting the coefficients before linearly independent functions $(u^2+1)\log u-u^2+1$, $u^2-1$, $2u,u^2+1$ we obtain the system of equations: $$C_6(m,n)=0,\quad C_6(m+1,n)=0,\quad C_2(m-1)+C_2(m)=0,$$ $$C_2(m-1)-C_2(m)+2C_5(m,n)=0.$$ Thus $$C_2(m)=-(-1)^mC,\quad C_6(m,n)=0,\quad C_5(m,n)=(-1)^m(-1)^nC.$$ Due to the last equalities we find the final form of the symmetry desired:
$$g=(-1)^{m+n}C\frac{u(u^2-1)(u_1u_{-1}+1)}{(uu_1+u_1-u+1)(uu_{-1}-u_{-1}+u+1)} $$

\section{Evaluation of a fifth order symmetry for quad graph equation}

In this section we proceed with the testing of our symmetry finding algorithm. Apply it again to equation (\ref{introductionHV}). Now we will look for a fifth order symmetry for it 
\begin{equation}\label{5sym}
u_t=g(u_{2},u_{1},u,u_{-1},u_{-2}).
\end{equation}
Start with the equations (\ref{Y1z}) and (\ref{Y-1z}) for the function $z=\log g_{u_2}$. These equations being as follows $Y_1z=f^{(1)},\ Y_{-1}z={f}^{(-1)}$ in an enlarged form are
\begin{eqnarray}
&&{\frac { ( u_{-1}^2-1 )( u_{-2}^2-1 )   }{4u_{-1}u}}\frac{\partial z}{\partial u_{-2}}+{\frac {  u_{-1}^2-1 }{2u
}}\frac{\partial z}{\partial u_{-1}}+\frac{\partial z}{\partial u}+\nonumber \\ 
&&+{\frac {2u_1}{  u^2-1}}\frac{\partial z}{\partial u_1}+
 {\frac {4u_2u_1}{ ( u_1^2-1 ) ( u^2-1 ) }}
\frac{\partial z}{\partial u_{2}}=  
\frac1{u}-\frac{4u_1}{(u_1^2-1)(u^2-1)};\label{Y1f1} \\ 
&&{\frac {4u_{-2}u_{-1}}{ ( u^2-1 )( u_{-1}^2-1 ) }}\frac{\partial z}{\partial u_{-2}}
-{\frac {2u_{-1}}{  u^2-1}}\frac{\partial z}{\partial u_{-1}}+\frac{\partial z}{\partial u}-\nonumber \\ 
&& -{\frac { u_1^2-1  }{2u
}}\frac{\partial z}{\partial u_1}
+{\frac { ( u_2^2-1 )  ( u_1^2-1 )   }{4u_1u}}
\frac{\partial z}{\partial u_2} =
{\frac {2u}{  u^2-1 }}-{\frac {  u_2 ( u_1^2-1 ) }{2u_1u}}.\label{Y-1f-1} 
\end{eqnarray}
Obviously system (\ref{Y1f1}), (\ref{Y-1f-1}) is not closed. To close it we add its differential consequences obtained by taking cross applications of the operators $Y_1$, $Y_{-1}$:
\begin{eqnarray}
&&Y_1z=f^{(1)},\nonumber\\
&&Y_{-1}z={f}^{(-1)},\nonumber\\
&&[Y_1,Y_{-1}]z=Y_1{f}^{(-1)}-Y_{-1}{f}^{(1)}=:\widetilde f,\label{closedsystem}\\
&&[Y_1,[Y_1,Y_{-1}]]z=Y_1\widetilde f-[Y_1,Y_{-1}]f^{(1)},\nonumber\\
&&[Y_{-1},[Y_1,Y_{-1}]]z=Y_{-1}\widetilde f-[Y_1,Y_{-1}]f^{(-1)}.\nonumber
\end{eqnarray}
System (\ref{closedsystem}) is closed and consistent. Solve it respect to the derivatives and find
$$\frac{\partial z}{\partial u_{-2}}=0,\quad \frac{\partial z}{\partial u_{-1}}=0,\quad \frac{\partial z}{\partial u_{2}}=-{\frac {2(u_{{1}}+1)}{1-u_{{1}}+u_{{2}}+u_{{2}}u_{{1}}}}, $$

$$\frac{\partial z}{\partial u}=\frac1{u}+\frac{2(1-u_1)}{ u_{{1}}+1+u_{{1}}u-u } + \frac{2u}{ u^2-1 },$$
$$\frac{\partial z}{\partial u_{1}}=\frac1{u_{{1}}}-{\frac {2(u+1)}{u_{{1}}+1+u_{{1}}u-u}}+\frac{2(1-u_2)}{ 1
-u_{{1}}+u_{{2}}+u_{{2}}u_{{1}}}+ \frac{2u_1}{ u_{{1}}^2-1 }.$$
Find $z$ by integrating:
$$z=\log   {\frac {u_{{1}}u \left( u_{{1}}^2-1 \right) \left( u^2-1 \right)  \left( u+1 \right) }{
 \left( 1-u_{{1}}+u_{{2}}+u_{{2}}u_{{1}} \right) ^{2} \left( u_{{1}}+1
+u_{{1}}u-u \right) ^{2}}}, 
$$ 
which yields immediately  an intermediate representation of $g$:
$$g={\frac {uu_{{1}} \left( u_{{1}}-1 \right)  \left( u^2-1
 \right)   }{ \left( u_{{1}}+1+u_{{1}}u-
u \right) ^{2} \left( 1-u_{{1}}+u_{{2}}+u_{{2}}u_{{1}} \right) }
}+g_1(u_{1},u,u_{-1},u_{-2})$$

For the further specification of $g$ let us find $v(u_{1},u,u_{-1},u_{-2})=\log g_{u_{-2}}=\log g_{1,u_{-2}}$. Function $v=v(u_{1},u,u_{-1},u_{-2})$ should satisfy equations (compare with (\ref{sv})):
\begin{eqnarray}
&&{\frac { ( u_{-1}^2-1 )( u_{-2}^2-1 )   }{4u_{-1}u}}\frac{\partial v}{\partial u_{-2}}+{\frac {  u_{-1}^2-1 }{2u
}}\frac{\partial v}{\partial u_{-1}}+\frac{\partial v}{\partial u}+\nonumber\\ 
&&+{\frac {2u_1}{  u^2-1}}\frac{\partial v}{\partial u_1}=
{\frac {2u}{ \left( u+1 \right)  \left( u-1
 \right) }}-{\frac { \left( u_{{-1}}+1 \right) u_{{-2}} \left( u_{{-1}}-1 \right) }{2u_{{-1}}u}}
;\label{sv11}\\
&&{\frac {4u_{-2}u_{-1}}{ ( u^2-1 )( u_{-1}^2-1 ) }}\frac{\partial z}{\partial u_{-2}}
-{\frac {2u_{-1}}{  u^2-1}}\frac{\partial v}{\partial u_{-1}}+\frac{\partial v}{\partial u}-\nonumber\\
&& -{\frac { u_1^2-1  }{2u
}}\frac{\partial v}{\partial u_1}
=\frac1{u}-{\frac {4u_{{-1}}}{ \left( u_{{-1}}+1 \right) 
 \left( u_{{-1}}-1 \right)  \left( u+1 \right)  \left( u-1
 \right) }}.\label{sv12}
\end{eqnarray}

System of equations (\ref{sv11}), (\ref{sv12}) can be closed by taking cross application of the operators $Y_1$ and $Y_{-1}$. After some manipulations of such a kind, which are omitted, one finds partial derivatives of v:
$$\frac{\partial v}{\partial u_{-2}}=\frac {2(1-u_{{-1}})}{1+u_{{-1}}-u_{{-2}}+u_{{-2}}u_{{-1}}},\quad \frac{\partial v}{\partial u_{1}}=0, $$
$$\frac{\partial v}{\partial u_{-1}}=\frac1{u_{{-1}}}-{\frac {2(u_{{-2}}+1)}{1+u_{{-1}}-u_{{-2}}+u_{{-2}}u_{{-
1}}}}-{\frac {2(u-1)}{-u_{{-1}}+u_{{-1}}u+1+u}}+\frac{2u_{-1}}{u_{-1}^2-1}
 $$
$$\frac{\partial v}{\partial u}=\frac1{u}-{\frac {2(u_{{-1}}+1)}{-u_{{-1}}+u_{{-1}}u+1+u
}}+\frac{2u}{u^2-1}.
$$
Thus $v$ is:
$$v=\log C{\frac {u_{{-1}}u   \left( u_{{-1}}^2-1
 \right)    \left( u^2-1 \right) }{
 \left( 1+u_{{-1}}-u_{{-2}}+u_{{-2}}u_{{-1}} \right) ^{2} \left( -u_{{
-1}}+u_{{-1}}u+1+u \right) ^{2}}}.$$
Since $v=\log g_{u_{-2}}=\log g_{1,u_{-2}}$ by integrating we obtain more detailed representation for $g$:
$$g={\frac {uu_{{1}} \left( u_{{1}}-1 \right)  \left( u^2-1
 \right)   }{ \left( u_{{1}}+1+u_{{1}}u-
u \right) ^{2} \left( 1-u_{{1}}+u_{{2}}+u_{{2}}u_{{1}} \right) }
}+$$ $$+C{\frac {u_{{-1}}u \left( u_{{-1}}+1 \right)   \left( u^2-1 \right) }{ \left( -u_{{-1}}+u_{{-1}}u
+1+u \right) ^{2} \left( 1+u_{{-1}}-u_{{-2}}+u_{{-2}}u_{{-1}}
 \right) }}+g_2(u_{-1},u,u_1),
$$
containing an unknown tail $g_2(u_{-1},u,u_1)$. Due to equation (\ref{y222}) function $y^{(1)}=\partial g_2/\partial u_1$ closely connected with $z^{(1)}=\partial g/\partial u_1$ solves the following system of equations
$${\frac { u_{{-1}}^2-1 }{2u}} {
\frac {\partial y^{(1)}}{\partial u_{{-1}}}}+
{\frac {\partial y^{(1)}}{\partial u}} +{\frac {2u_{{1}}}{ u^2-1 }}{\frac 
{\partial y^{(1)}}{\partial u_{{1}}}}-{
\frac { \left( {u}^{2}-1-2\,u \right) }{u ( u^2-1) }}{y^{(1)}} =$$ $${\frac { \left( u_{{-1}}^2+1 \right) 
 \left( u^2-1 \right)   u}{
 \left( u_{{1}}+1+uu_{{1}}-u \right) ^{2} \left( -u_{{-1}}
+1+uu_{{-1}}+u \right) ^{2}}};$$
$$
{\frac {2u_{{-1}}}{ u^2-1 }}{\frac 
{\partial y^{(1)}}{\partial u_{{-1}}}}-{\frac {\partial y^{(1)}}{\partial u}}+
{\frac { u_{{1}}^2-1 }{2u}} {\frac {\partial y^{(1)}}{\partial u_{{1}}}}+{\frac {2\,{u}^{2}+{u}^{2}u_{{1}}-u_{{1}}}{u \left( u^2-1 \right) }}y^{(1)}=
$$ $$\frac{u(u_1-1)^2}{u+1}\left(\frac2{(u_1+1+uu_1-u)^3}+{\frac {-2\,{u}^{2}u_{{-1}}-u+3\,u{u_{{-1}}}^{2}+2\,
u_{{-1}}-1-{u_{{-1}}}^{2}}{ ( -u_{{-1}}+1+uu_{{-1}}+u
 ) ^{2} ( u_{{1}}+1+uu_{{1}}-u ) ^{2}}}
\right).$$ 
These equations imply
$$\frac{\partial y^{(1)}}{\partial u_{-1}}={\frac {4{u}^{2} ( u^2-
1 ) ^{2} \left( u_{{1}}u_{{-1}}+1 \right) }{ \left( -u_{{-1}}+1+
uu_{{-1}}+u \right) ^{3} \left( u_{{1}}+1+uu_{{1}}-u
 \right) ^{3}}};$$
$$\frac{\partial y^{(1)}}{\partial u_{1}}+\frac {2 \left( u+1 \right) } {u_{{1}}+1+uu_{{1}}-u}y^{(1)}=-{\frac {2
 \left( u_{{-1}}-1 \right)  ( u^2-1 ) ^{2}u}{ \left( u_{{1}}+1+uu_{{1}}-u
 \right) ^{4} \left( -u_{{-1}}+1+uu_{{-1}}+u \right) }}.
$$ We omit the expression for $\frac{\partial y^{(1)}}{\partial u}$ since it is huge.

Thus we obtain the next representation for the searched $g$:
$$g={\frac {uu_{{1}} \left( u_{{1}}-1 \right)  \left( u^2-1
 \right)   }{ \left( u_{{1}}+1+u_{{1}}u-
u \right) ^{2} \left( 1-u_{{1}}+u_{{2}}+u_{{2}}u_{{1}} \right) }
}+$$ $$+C{\frac {u_{{-1}}u \left( u_{{-1}}+1 \right)   \left( u^2-1 \right) }{ \left( -u_{{-1}}+u_{{-1}}u
+1+u \right) ^{2} \left( 1+u_{{-1}}-u_{{-2}}+u_{{-2}}u_{{-1}}
 \right) }}+
$$ $${\frac { \left(-u_{{1}} \left( u^2-1 \right)  {u_{{-1}}}
^{2}+2\, \left( u-1 \right)  \left( uu_{{1}}-u+u_{{1
}} \right) u_{{-1}}+u_{{1}} \left( u+1 \right) ^{2}
\right)  \left( u-1 \right) u}{ \left( -u_{{-1}}+1+
uu_{{-1}}+u \right) ^{2} \left( u_{{1}}+1+uu_{{1}}-u
 \right) ^{2}}}- $$ $${\frac {{ C_2}\, \left( u-1 \right) u_{{0
}}}{u_{{1}}+1+uu_{{1}}-u}}+g_3(u_{-1},u),$$ where $C_2$ is the constant of integration. By substituting this repsentation of $g$ into linearized equation (\ref{linearisation}) and differentiating it with respect to $u_2$ we get $C_2=0$. In order to look for  $g_3$ differentiate the linearized equation respect to $u_{-1}$ and study according to the scheme above the linear equations for $y^{(2)}=\frac{\partial g_3}{\partial u_{-1}}$. Their consistency condition immediately implies $C=-1$. For this value of $C$ the equations mentioned are of the form:
$$\frac {  u_{{-1}}^2-1}  {2u}\frac{\partial y^{(2)}}{\partial u_{-1}}
+\frac{\partial y^{(2)}}{\partial u} +{\frac {  {u}^{2}u_{{-1}}-2\,{u}^{2}-u_{{-1}}   }{ \left( u^2-1 \right)   u
}}y^{(2)}=-{\frac { \left( 3\,uu_{{-1}}+u+u_{{-1}}+1 \right) 
 \left( u-1 \right) }{ \left( u+1 \right)  \left( 1-u_{{-1}
}+uu_{{-1}}+u \right) ^{3}}}$$

$${\frac { 2 u_{{-1}}}{ u^2-1  }}\frac{\partial y^{(2)}}{\partial u_{-1}}-\frac{\partial y^{(2)}}{\partial u} +{\frac { \left( {u
}^{2}-1+2\,u \right)  }{
 \left( u-1 \right)  \left( u+1 \right) u}}y^{(2)}={\frac {
 \left( u-1 \right)  \left( 1+u+u_{{-1}}-uu_{{-1}}
 \right) u}{ \left( u+1 \right)  \left( 1-u_{{-1}}+uu_{{-1}}+u \right) ^{3}}}
 $$
By solving these equations we find the next specification of $g$:
$$g={\frac {uu_{{1}} \left( u_{{1}}-1 \right)  \left( u^2-1
 \right)   }{ \left( u_{{1}}+1+u_{{1}}u-
u \right) ^{2} \left( 1-u_{{1}}+u_{{2}}+u_{{2}}u_{{1}} \right) }
}+$$ $$-{\frac {u_{{-1}}u \left( u_{{-1}}+1 \right)   \left( u^2-1 \right) }{ \left( -u_{{-1}}+u_{{-1}}u
+1+u \right) ^{2} \left( 1+u_{{-1}}-u_{{-2}}+u_{{-2}}u_{{-1}}
 \right) }}-
$$ 
$$ -{\frac {u \left( u^2-1 \right)  
 \left( 1+u_{{1}}u_{{-1}} \right)  \left( uu_{{-1}}-u_{{-1}}-uu_{{1}}-u_{{1}} \right) }{ \left( u_{{1}}+1+uu_{{1}}-u
 \right) ^{2} \left( 1-u_{{-1}}+uu_{{-1}}+u \right) ^{2}}
}-{\frac { \left( u+1 \right) u{C_3}}{-u_{{-1}}+1+uu_{{-1}}+u}}+ g_4 ( u ).$$
By substituting it into linearized equation and differentiating the result with respect to the variable $u_{-1}$ we prove that  $C_3=0$ and that  $g_4(u)$ solves a linear homogeneous equation which has unique solution $g_4=0$. Thus the final form of the symmetry searched is (see also (\ref{introductionHVsym}) above)
$$u_t=g={\frac {uu_{{1}} \left( u_{{1}}-1 \right)  \left( u^2-1
 \right)   }{ \left( u_{{1}}+1+u_{{1}}u-
u \right) ^{2} \left( 1-u_{{1}}+u_{{2}}+u_{{2}}u_{{1}} \right) }
}+$$ $$-{\frac {u_{{-1}}u \left( u_{{-1}}+1 \right)   \left( u^2-1 \right) }{ \left( -u_{{-1}}+u_{{-1}}u
+1+u \right) ^{2} \left( 1+u_{{-1}}-u_{{-2}}+u_{{-2}}u_{{-1}}
 \right) }}-
$$ 
$$ -{\frac {u \left( u^2-1 \right)  
 \left( 1+u_{{1}}u_{{-1}} \right)  \left( uu_{{-1}}-u_{{-1}}-u_{
0}u_{{1}}-u_{{1}} \right) }{ \left( u_{{1}}+1+uu_{{1}}-u
 \right) ^{2} \left( 1-u_{{-1}}+uu_{{-1}}+u \right) ^{2}}
}.$$

\section{Higher order characteristic vector fields and additional differential consequences of integrability conditions}

For some certain cases systems (\ref{s}), (\ref{sv}), (\ref{econd3}), (\ref{econd4}) derived in section 3 are sufficient to find the final form of the symmetry searched if it does exist. But for the other ones these systems do not provide enough information to find the symmetry desired or to make the final decision on its existence. Here we briefly discuss how to derive additional systems of linear first order PDE consistency of which is also necessary for existence of higher symmetries. To this end we can make of use the higher characteristic vector fields defined in such a way 
\begin{equation}\label{hvf}
Y_k=\bar D^{-k}X_{-1}\bar D^{k}\quad \mbox{and} \quad Y_{-k}=\bar D^{k}X_{1}\bar D^{-k}\quad \mbox{for}\quad \forall k>1. 
\end{equation}
It follows from equation (\ref{z}) that $\bar z_2=z+(1+\bar D)(D^{-1}-D^{j-1})\log f_{u_1}$. By applying the operator $\bar D^{-2}X_{-1}$ to the last equation we find
\begin{equation}\label{Y2z}
Y_{2}z=f^{(2)},
\end{equation}
where $f^{(2)}=\bar D^{-2}\frac{\partial}{\partial \bar u_1}(1+\bar D)(D^{-1}-D^{j-1})\log f_{u_1}$.
The coordinate representation of the vector field $Y_2$ is as follows
\begin{equation}\label{defY2}
Y_2=X_1+\bar D^{-1}(Y_1f)\frac{\partial}{\partial u_1}+\bar D^{-1}(Y_1f^{-1,1})\frac{\partial}{\partial u_{-1}}+ \bar D^{-1}(Y_1f_1)\frac{\partial}{\partial  u_{2}}+\bar D^{-1}(Y_1f^{-1,1}_{-1})\frac{\partial}{\partial  u_{-2}}+ ...\,.
\end{equation}
The coefficients of this equation depend on the extra variable $\bar u_{-2}$, hence we have to require
\begin{equation}\label{X2}
X_{2}z=0,
\end{equation}
where $X_2=\frac{\partial}{\partial \bar u_{-2}}$. 

Represent now equation (\ref{z}) as $\bar z_{-2}=z-(\bar D^{-1}+\bar D^{-2})(D^{-1}-D^{j-1})\log f_{u_1}$. By applying the reasonings above one obtains a pair of equations 
\begin{equation}\label{Y-2z}
Y_{-2}z=f^{(-2)},\quad X_{-2}z=0
\end{equation}
where $f^{(-2)}=-Y_{-2}(1+\bar D)(D^{-1}-D^{j-1})\log f_{u_1}$ and $X_{-2}=\frac{\partial}{\partial \bar u_{2}}$.

Iterating this way one obtains a system of the form
\begin{eqnarray}\label{si}
Y_{i}z&=&f^{(i)},\\
X_{i}z&=&0, \qquad -N\leq i\leq N,\quad i\neq0,\nonumber
\end{eqnarray}
which should be consistent for any value of the natural $N$. Here $X_i=\frac{\partial}{\partial \bar u_{-i}}$, functions $f^{(i)}$ are expressed through the given functions $f,f^{1,-1},f^{-1,1},f^{-1,-1}$ and their derivatives and shifts. Recall that functions $f^{(1)}$, $f^{(-1)}$ are defined earlier in (\ref{Y1z}), (\ref{Y-1z}).

Obviously such kind extended consequences can be deduced from any integrability condition discussed in \S 2.

\section{Conclusions}

Integrability conditions for quad graph equations are studied based on the symmetry approach. Effective consequences of existence of higher symmetries are derived by using characteristic vector fields.  These conditions can be successfully used for testing and for classification of integrable quad graph equations. An algorithm of evaluating higher symmetries (generally non-autonomous) for autonomous quad graph equations is suggested.  Efficiency of the algorithm is approved by applying to an example recently found  in \cite{HydonViallet}. Fifth and third order symmetries for this equation are evaluated, the latter turns out to be non-autonomous. We have to emphasize that the main difficulty of non-autonomous case is the circumstance  that all of the constants of integration, appearing when we solve the corresponding systems of PDE should now depend on the discrete arguments $m,n$.

\section*{Acknowledgments}
This work is partially supported by Russian Foundation for Basic
Research (RFBR) grants $\#$ 10-01-91222-CT-a, $\#$ 11-01-97005-r-povoljie-a, and $\#$ 10-01-00088-a.

\end{document}